# Critical behavior of slider-block model

## (Short title: Critical…)

### S G Abaimov

E-mail: sgabaimov@gmail.com.

**Abstract.** This paper applies the theory of continuous phase transitions of statistical mechanics to a slider-block model. The slider-block model is chosen as a representative of systems with avalanches. Similar behavior can be observed in a forest-fire model and a sand-pile model. Utilizing the well-developed theory of critical phenomena for percolating systems as a foundation, a strong analogy for the slider-block model is developed. It is found that the slider-block model has a critical point when the stiffness of the model is infinite. Critical exponents are found and it is shown that the behavior of the slider-block model and, particularly, the occurrence of system-wide events are strongly dominated by finite-size effects. Also the unknown before behavior of the frequency-size distributions is found for large statistics of events.

## 1. Introduction

Models with avalanches, recently introduced in the literature, exhibit complex behavior of event occurrence. A slider-block model [1] (further on SBM) has been investigated by many studies as a model representing the recurrent earthquake occurrence [2-5]. A forest-fire model represents the occurrence of fires in forests [6, 7]. A sand-pile model [8] is the main representative of the theory of self-organized criticality. All these models exhibit complex behavior. Energy (or another driving quantity) is pumped into a system. In response the system organizes its dissipation through the complex behavior of avalanches.

During the past century major breakthroughs have been achieved in the theory of phase transitions in statistical mechanics (for reviews see, *e.g.* [9-12]). The major concepts of this theory have been applied not only to the typical thermal systems like liquid-gas or magnetic systems but also to systems without actual termalization like percolation theory [13] or damage mechanics [14, 15]. In this paper we apply the concepts of continuous phase transitions to the slider-block model. For the forest-fire and sand-pile models the application of statistical mechanics is similar and will be investigated in future publications.

Grassberger P. [7] has shown that critical exponents significantly depend on the model size. In this paper we investigate this effect for the slider-block model. We consider models with $L = 25$, 50, 100, 500, and 1000 slider-blocks. Also in preliminary studies we discovered that the numbers of events in statistics also significantly influence the critical behavior.

Particularly, we found that the dependence of a correlation length on a tuning field parameter exhibits significant non-smooth deviations for statistics of 10,000 events in comparison with statistics of 1,000,000 events which we used as a reference. The frequency-size behavior also significantly depends on the size of statistics. We found unknown behavior when we used large statistics.

We utilize the modification of the SBM which requires integration of coupled ordinary differential equations. Therefore the sizes of statistics are limited by the time of numerical simulations. In spite of this difficulty for large model sizes of $L = 500$ and $L = 1000$ blocks we have obtained large statistics in the range from 170,000 up to 1,100,000 avalanches. Most of distributions have from 300,000 to 800,000 slip events. This lets us obtain smooth scaling dependences and accurate values of critical exponents.

In Section 2 we introduce the model. In Section 3 we investigate its frequency-size behavior. In Section 4 we consider an analogy with the theory of percolation and develop preliminary expectations what a critical point and a correlation length of the model are. In Section 5 we develop a rigorous expression for the correlation length and consider its behavior. Also we investigate the finite-size scaling of the model and find that the dependences of correlation length for different model sizes collapse on a single curve, representing a scaling function. In Sections 6 we investigate the scaling behavior of a susceptibility and also find its scaling function. In Section 7 we return to the frequency-size distribution and investigate its scaling behavior. Although all correlation length, susceptibility, and frequency-size distribution represent correlations of fluctuations, the main representative is a correlation function. In Section 8 we investigate its scaling behavior.

## 2. The model

In this paper we utilize a modification of the slider-block model (SBM) with the inertia of blocks where the differential equations of motion are coupled [2]. This is the variation of the model which is the most difficult to simulate numerically. However, it has an advantage of the absence of multiple approximations that are used in other modifications. One of the most important improvements is that the time evolution of an avalanche includes coupled motion of all participating blocks in contrast to cellular-automata models where blocks move in sequences (i.e., a block can move only when its neighbor stops).

A linear chain of $L$ slider blocks of mass $m$ is pulled over a surface at a constant velocity $V_L$ by a loader as illustrated in figure 1. This introduces a mechanism to pump energy into the system. Each block is connected to the loader by a spring with stiffness $k_L$. Adjacent blocks are connected to each other by springs with stiffness $k_C$. Boundary conditions are assumed to be periodic: the last block is connected to the first block.

The blocks interact with the surface through static-dynamic friction. The static stability of each slider-block is given by

$$k_L y_i + k_C \left( 2 y_i - y_{i-1} - y_{i+1} \right) < F_{Si} , \tag{1}$$

where $F_{Si}$ is the maximum static friction force on block $i$ holding it motionless and $y_i$ is the position of block $i$ relative to the loader. These thresholds introduce the non-linearity of system's behavior.

During strain accumulation due to the loader motion all blocks are motionless relative to the surface and have the same increase of their coordinates relative to the loader plate

$$\frac{dy_i}{dt} = V_L . \tag{2}$$

When the cumulative force of the springs connecting to block $i$ exceeds the maximum static friction $F_{Si}$, the block begins to slide. The dynamic slip of block $i$ is controlled by its inertia

$$m \frac{d^2 y_i}{dt^2} + k_L y_i + k_C \left( 2 y_i - y_{i-1} - y_{i+1} \right) = F_{Di} , \tag{3}$$

where $F_{Di}$ is the dynamic (sliding) frictional force on block $i$. The loader velocity is assumed to be much smaller than the slip velocity, so the movement of the loader is neglected during a slip event. This is consistent with the concept that the slip duration of an earthquake is negligible in comparison with the interval of slow tectonic stress accumulation between earthquakes.

The sliding of one block can trigger instability of other blocks forming a multi-block event. When the velocity of a block decreases to zero it sticks and switches from the dynamic to static friction.

It is convenient to introduce the non-dimensional variables and parameters: $\tau_f = t\sqrt{\dfrac{k_L}{m}}$ for the fast time during avalanche evolutions, $Y_i = \dfrac{k_L y_i}{F_S^{ref}}$ for the coordinates of blocks. The ratio of static to dynamic friction $\phi = \dfrac{F_{Si}}{F_{Di}}$ is assumed to be the same for all blocks $\phi = 1.5$ but the values of friction $\beta_i = \dfrac{F_{Si}}{F_S^{ref}}$ vary from block to block with $F_S^{ref}$ as a reference value of the static frictional force ($F_S^{ref}$ is the minimum value of all $F_{Si}$). Particularly, the values of frictional parameters $\beta_i$ are assigned to blocks by the uniform random distribution in the range $1 < \beta_i < 3.5$. This quenched random disorder in the system is a 'noise' required to generate event's variability in stiff systems. Parameter $\alpha = \dfrac{k_C}{k_L}$ is the stiffness of the system relative to the stiffness of system's connection to the loader. Later we will see that $\alpha$ plays an important role of a tuning field parameter. For all model sizes as values of $\alpha$ we will in general utilize 1, 2, 3, 4, 5, 6, 7, 8, 9, 10, 11, 12, 14, 16, 18, 20, 25, 30, 35, 40, 50, 60, 75, 100, 200, 500, 1000, 2000, 5000, 10000, and some others, specific for each particular model size.

Stress accumulation occurs when all blocks are stable; slip of blocks occurs during the fast time $\tau_f$ when the loader is assumed to be approximately motionless. In terms of these non-dimensional variables the static stability condition (1) becomes

$$Y_i + \alpha(2Y_i - Y_{i-1} - Y_{i+1}) < \beta_i, \tag{4}$$

strain accumulation (2) becomes

$$\frac{dY_i}{d\tau_s} = 1, \tag{5}$$

and dynamical slip (3) becomes

$$\frac{d^2 Y_i}{d\tau_f^2} + Y_i + \alpha(2Y_i - Y_{i-1} - Y_{i+1}) = \frac{\beta_i}{\phi}. \tag{6}$$

For numerical simulations a velocity-verlet numerical scheme is utilized which is a typical scheme for molecular-dynamics simulations [e.g., 16].

## 3. Frequency-size behavior

Figures 2(a-b) illustrate behavior of the SBMs consisting of $L = 500$ and 1000 blocks. The probability density function of the frequency-size distribution is plotted on log-log axes for different values of the system stiffness $\alpha$. As a size $s$ of an event the number of different blocks participating in this event is used. During an avalanche a block can lose and gain its stability many times but is counted only once in the size of this event. This makes the size of an event equal to its elongation in the model space (equal to the number of consecutive blocks in a continuous chain which has lost its stability). If the size of an event equals the size of the model we will refer to these events as system-wide (SW) events. For large sizes in figures 2(a-b) the sliding average over 9 adjacent sizes has been used to remove fluctuations.

For small values of $\alpha$ the SBM has no SW events. The frequency-size statistics for small events has a tendency to be similar to the Gutenberg-Richter power-law distribution (straight line on the log-log axes) but for larger events it has a roll-down. When $\alpha$ increases, the roll-down moves to the right and finally goes beyond the system size $L$. Also the behavior of the system changes: We see the appearance of a peak of events whose sizes are about a half of the model size. When $\alpha$ exceeds some critical value, the first SW events start to appear. The peak of the half-model-size events becomes narrower and disappears on some statistics ( $\alpha = 2000$ for $L = 500$ blocks and $\alpha = 5000$ for $L = 1000$ blocks). Instead, another peak appears which is adjacent to the SW limit. We believe that this effect is observed for the first time in this study due to the presence of large statistics.

Further increase of $\alpha$ is assumed to cause all complexities of the curve to disappear and the frequency-size dependence is assumed to become a perfect power-law plus a discrete peak of SW events. However, we have not been able to observe this effect for the large systems with $L = 500$ and $L = 1000$ blocks because this clean power-law dependence is supposed to appear at very high values of $\alpha$, where the differential equations become difficult to be solved numerically. Therefore we illustrate this dependence for the model with $L = 100$ blocks in figure 2(c). The maximum likelihood fit gives the value of the exponent of the power-law dependence $\tau = 2.08 \pm 0.09$ which is very close to 2. Therefore we can suggest that in the limit of infinite stiffness the model exhibits the power-law dependence of non-SW events with the meanfield (rational) value $\tau = 2$ of the exponent. We illustrate this model tendency in figure 3. The frequency-size distributions are normalized by the number of SW events. For all model sizes we used here the same value of $\alpha = 1000$. When the size of the model decreases we see the tendency of the distribution to attenuate the peak of half-model-size events and to become a power-law plus the discrete peak of SW events. We see that $\alpha = 1000$ is sufficient to reveal the power-law tendency for model sizes $L = 25$, 50, and 100. However, for model sizes $L = 500$ and 1000 the system is not stiff enough to remove the influence of the peak of half-model-size events from the power-law dependence.

Also in figure 3 we see that for the same event size $s$ the number of events with this size relative to the number of SW events increases with the increase of the model size. However, this increase is less than an order of amplitude and can be associated with the deviations from the pure power-law dependence. Again, these deviations are caused by the fact that the stiffness $\alpha$ of the system is not high enough.

The dependence of $\alpha$, at which the first SW events appear, on the size of the model is shown in figure 4. We see that the appearance of the first SW events depends on the system size and is a result of the finite-size effect [3]. Therefore, it would be wrong to interpret the appearance of the first SW events in a finite system as a critical point of the infinite model. What the meaning of these values of $\alpha$ is and what the critical point of the model is, we will discuss in the next section.

## 4. An analogy with the percolation theory

As a possible analogy we consider a percolating system. In the case of site percolation [13] a field parameter $p$ is the probability for a lattice site to be occupied. If $N_\bullet$ is the number of occupied sites on the lattice and $N_{\text{total}}$ is the total number of sites on the lattice then $p = N_\bullet / N_{\text{total}}$. For the rectangular (square) $d$-dimensional lattice with the linear size of $L$ sites the total number of sites is $N_{\text{total}} = L^d$.

For the given value of $p$ we define a microstate as a particular microconfiguration of occupied sites realized on the lattice. For example, for $N_\bullet = 1$ there are $N_{\text{total}}$ microstates when there is only one occupied site at any of $N_{\text{total}}$ possible locations on the lattice. For $N_\bullet = N_{\text{total}}$ there is only one microstate when all sites are occupied.

Let us assume that $p$ increases from 0 to 1. Then initially for $p$ below the percolation threshold $p_C$ there is no percolating cluster on the infinite lattice. For the finite lattice with size $L$ for $p < L / N_{\text{total}}$ (for $N_\bullet < L$) there is also no percolating cluster. However, when $p$ is greater than $L / N_{\text{total}}$ (when $N_\bullet \geq L$) the appearance of a percolating cluster among all microstates is possible. Particularly, percolating is any microstate which contains one row of the lattice completely occupied. For $p$ significantly below the percolating threshold $p_C$ the number of these percolating microstates is much smaller than the total number of microstates for the given $p$. Therefore if an observer were looking at an ensemble of all possible system's realizations for the given $p$ (an ensemble of all possible microstates) s/he would count percolating microstates as highly improbable and their fraction among all microstates in the ensemble as negligible. A correlation length $\xi$ for this value of $p$ is much smaller than the system size.

Even when, for the further increase of $p$, the fraction of percolating microstates becomes finite for a finite system, this does not guarantee that the infinite system percolates. In the finite system the fraction of percolating microstates becomes finite earlier than in the infinite system because of the finite-size effect [13]. The finite system begins to percolate when the correlation length $\xi$ reaches the size of the system $L$. But percolation of the infinite system requires an infinite correlation length (which appears at higher values of $p$). Visually, the infinite system can be imagined as composed by an ensemble of finite systems combined together (an ensemble of all microstates of a finite system). If only the negligible fraction of these microstates percolate the finite lattice, then the infinite system does not have a percolating cluster.

For the case of the SBM the field parameter is the stiffness of the system $\alpha$. For small values of $\alpha$ there are no SW events in the system. If $\alpha$ increases and exceeds some threshold, the first SW events appear in the system. However, the fraction of these events (*e.g.,* 3 of 290,000 for the system size $L = 1000$ and $\alpha = 14$) is very small. The appearance of SW events is possible because the field parameter is above some threshold. However, the correlation length $\xi$ is still finite and, in fact, is very much smaller than the size of the system $L$. For further increase of the field parameter the correlation length reaches the size of the finite system, but is still much smaller than the infinite correlation length, required for the infinite system to reach its critical point. Therefore, the first appearance of SW events in the finite system must not be confused with the case of an infinite system at the critical point $\alpha_C$.

Returning again to the percolating system, below the percolation threshold $p < p_C$ the behavior of the system is significantly different at different spatial scales. For the scales smaller than the correlation length $\xi$ the distribution of clusters is fractal and scale-invariant. The frequency-size distribution of cluster sizes in this case is a power-law and again there is an analogy here with the frequency-size distribution of small events in the SBM (straight line of the Gutenberg-Richter power-law distribution for small events on log-log axes). For the scales similar or greater than the correlation length $\xi$ the frequency-size distribution of clusters deviates from the power-law and has an exponential roll-down. Again, there is an analogous roll-down for the SBM for larger events. Therefore, preliminary, for the SBM the correlation length $\xi$ can visually be found as being in the range where the frequency-size distribution changes its behavior from the power-law to the roll-down. However, the frequency-size behavior of the SBM is more complex than the behavior of the percolating system. Therefore later we will provide a more rigorous statement.

For a finite percolating system, when the correlation length becomes greater than the system size $\xi \geq L$, the distribution of all non-percolating clusters is fractal and scale-invariant. The same we can see for the SBM for the range of high values of $\alpha$ when the roll-down has moved completely beyond the system size $L$ and the frequency-size distribution for non-SW events becomes a power-law distribution $\text{pdf}(S) \propto S^{-2}$ ($\alpha = 1000$ in Fig. 2(c)). When the correlation length approaches the system size for a finite system, the fraction of percolating clusters becomes finite because of the finite-size effect. Therefore for the SBM we can conclude that the appearance of the significant fraction of SW events is also a result of the finite-size effect and is an indication that the correlation length $\xi$ is reaching the system size $L$.

## 5. Correlation length $\xi$

First we will consider the definition of a correlation length in the theory of percolation. For the infinite system the correlation length $\xi$ may be defined as the averaged root mean square distance between two arbitrary occupied sites on the lattice under the condition that these two sites must belong to the same cluster [13]

$$\xi \equiv \sqrt{\frac{\sum\limits_{<i,j>\in \text{ the same cluster}} r_{i,j}^2}{\sum\limits_{<i,j>\in \text{ the same cluster}} 1}},\qquad(7)$$

where indexes $i$ and $j$ enumerate occupied sites on the lattice, $r_{i,j}$ is the distance between occupied sites $i$ and $j$, and sum $\sum\limits_{<i,j>\in \text{ the same cluster}}$ goes over all pairs of occupied sites $<i,j>$ under the condition that both sites in each pair must belong to the same cluster. This definition can be written as averaging over all clusters on the lattice

$$\xi = \sqrt{\frac{\sum\limits_{k}\sum\limits_{<i,j>\in \text{ cluster } k} r_{i,j}^2}{\sum\limits_{k}\sum\limits_{<i,j>\in \text{ cluster } k} 1}},\qquad(8)$$

where index $k$ enumerates all clusters on the lattice. The sum over all clusters $\sum\limits_{k}$ can be transformed into the sums over different cluster sizes $s$ ($s$ is the number of occupied sites in a cluster)

$$\xi = \sqrt{\frac{\sum_s \sum_{k_s=1}^{N_s} \sum_{<i,j> \in \text{cluster } k_s} r_{i,j}^2}{\sum_s \sum_{k_s=1}^{N_s} \sum_{<i,j> \in \text{cluster } k_s} 1}} = \sqrt{\frac{\sum_s \sum_{k_s=1}^{N_s} \sum_{<i,j> \in \text{cluster } k_s} r_{i,j}^2}{\sum_s \sum_{k_s=1}^{N_s} \frac{1}{2}s(s-1)}} = \sqrt{\frac{\sum_s \sum_{k_s=1}^{N_s} \sum_{<i,j> \in \text{cluster } k_s} r_{i,j}^2}{\sum_s N_s \frac{1}{2}s(s-1)}},$$ (9)

where index $k_s$ enumerates $N_s$ clusters of size $s$. The radius of gyration of given cluster $k_s$ is

$$R_{k_s} = \sqrt{\frac{\sum_{<i,j> \in \text{cluster } k_s} r_{i,j}^2}{\sum_{<i,j> \in \text{cluster } k_s} 1}} = \sqrt{\frac{\sum_{<i,j> \in \text{cluster } k_s} r_{i,j}^2}{\frac{1}{2}s(s-1)}},$$ (10)

and the averaged root mean square radius of gyration for clusters of size $s$ on the lattice is

$$R_s = \sqrt{\frac{\sum_{k_s=1}^{N_s} R_{k_s}^2}{N_s}} = \sqrt{\frac{\sum_{k_s=1}^{N_s} \sum_{<i,j> \in \text{cluster } k_s} r_{i,j}^2}{N_s \frac{1}{2}s(s-1)}}.$$ (11)

Therefore equation (9) can be written as

$$\xi = \sqrt{\frac{\sum_s N_s \frac{1}{2}s(s-1)R_s^2}{\sum_s N_s \frac{1}{2}s(s-1)}} = \sqrt{\frac{\sum_s n_s \; s(s-1)R_s^2}{\sum_s n_s \; s(s-1)}},$$ (12)

where $n_s$ is the number of clusters of size $s$ per lattice site for the given $p$. So, the correlation length is the root mean square of radii of all clusters averaged over all clusters in the lattice not directly but with the weight coefficients $s(s-1)$.

Because the SBM is a one-dimensional chain of blocks, each event is assumed to be continuous over the model space (all blocks, which are unstable during an avalanche, form a continuous chain). Therefore for the SBM the size $s$ of an event (the number of blocks participating in an avalanche) is the elongation of this event. This significantly simplifies all further calculations. For any event of size $s$ the first site makes $(s-1)$ pairs with $(s-1)$ other sites. Then the second site makes $(s-2)$ pairs with $(s-2)$ sites, and so on. Finally, the site before the last site makes one pair with the last site. For the radius of gyration of this event it provides

$$R_{k_s} = \sqrt{\frac{\sum_{i=1}^{s-1} i^2(s-i)}{\sum_{i=1}^{s-1} i}} = \sqrt{\frac{\frac{1}{12}(s-1)s^2(s+1)}{\frac{1}{2}(s-1)s}} \ . \tag{13}$$

In the one-dimensional case for the same size $s$ there is no variability of clusters. Therefore the averaged radius of gyration $R_s$ equals to the radius of gyration of any cluster with size $s$: $R_s = R_{k_s}$.

For the correlation length $\xi$ in the similar way we obtain

$$\xi = \sqrt{\frac{\sum_{s=1}^{L} \mathrm{pdf}(s) \sum_{i=1}^{s-1} i^2(s-i)}{\sum_{s=1}^{L} \mathrm{pdf}(s) \sum_{i=1}^{s-1} i}} = \sqrt{\frac{\sum_{s=1}^{L} \mathrm{pdf}(s) \frac{1}{12}(s-1)s^2(s+1)}{\sum_{s=1}^{L} \mathrm{pdf}(s) \frac{1}{2}(s-1)s}} \ , \tag{14}$$

where pdf($s$) is the probability density function to observe an event with the elongation $s$ in the sequence of avalanches.

Figure 5(a) presents the dependence of the correlation length $\xi$ on the system stiffness $\alpha$ for different values of the model size $L$. Behavior of the correlation length suggests that the critical point is located in the infinity of the field parameter $\alpha$. Therefore further on we use a field parameter $t = 1 / \alpha$ instead of $\alpha$ and assume that the critical point is located at $t = 0$. Figure 5(b) presents on log-log axes the dependence of the correlation length $\xi$ on the field parameter $t$ for different model sizes. The correlation length $\xi$ increases monotonically with the decrease of $t$. Initially this increase is influenced by non-linear effects because the system is far from the fixed point of a renormalization group. When the field parameter $t$ reaches the vicinity of the fixed point, the linearization of the renormalization group becomes possible. Starting from this value of $t$ the dependence of the correlation length on the field parameter becomes a power-law $1/t^{\nu}$ with the exponent $\nu = 1.85 \pm 0.03$. This value was obtained by the maximum likelihood fit of the power-law parts of the curves for the SBMs with 500 and 1000 blocks. We use for the fit only these model sizes because they provide the dependence which is the cleanest from the non-linear and crossover effects. For the infinite system we would expect the power-law divergence of the correlation length at the critical point $t = 0$ with the same value of the exponent $\nu$.

However, our SBMs are finite. Therefore, for further decrease of $t$ the correlation length increases as a power-law and finally becomes of the order of the system size $L$. Starting from this value of the field parameter, the finite-size effect, as a crossover effect, influences the dependence of $\xi$ on $t$. When the system approaches the critical point, the correlation length reaches the limit of the system size and stays constant at this limit [9]. More rigorously, the averaged cluster elongation reaches the system size while the correlation length stays constant at a lower value due to the fact that correlation length (7) is always lower than the averaged linear size of clusters.

From the theory of scaling functions [9, 11, 17] we expect that the functional dependence of the correlation length $\xi$ on the field parameter $t$ should have the form

$$\xi \propto \frac{1}{t^\nu} \Xi(Lt^\nu), \text{ where } \Xi(x) = \begin{Bmatrix} const, \, x >> 1 \\ x, \, x << 1 \end{Bmatrix} \qquad (15)$$

is some scaling function. In the limit $Lt^\nu >> 1$, when $\xi << L$, this function has a constant limit, which does not influence the power-law dependence $1/t^\nu$. In the limit $Lt^\nu << 1$, when the correlation length of the infinite system is $\xi_\infty >> L$, this function generates a power-law dependence $x \propto t^\nu$, which cancels the power-law dependence $1/t^\nu$ in front of the function $\Xi(x)$ and provides the finite limit for the correlation length. To find the scaling function $\Xi(x)$, we multiply the dependence $\xi(t)$ by $t^{\nu}$ and then plot the resulting dependence $\xi t^{\nu}$ as a function of the parameter $x = Lt^{\nu}$. The obtained scaling function $\Xi(x)$ is presented in figure 6. Also we plot the dependence $x$ to compare it with the scaling function for low values of $x$. In figure 6 we see that all curves perfectly collapse on the scaling dependence $\Xi(x)$ except only for high values of $t$ when the system is far from the critical point and the renormalization group cannot be linearized far from its fixed point. At these high values of $t$ the power-law divergence $1/t^{\nu}$ of the correlation length has non-linear deviations, and the scaling is not valid.

## 6. Susceptibility *K*

Similarly to the correlation length, in this section we investigate the behavior of the susceptibility as a measure of fluctuations. In statistical mechanics this quantity is proportional to the variance of fluctuations; in the theory of percolation this quantity is called a mean cluster size [13]. Following the analogy with the percolation theory, we define susceptibility as

$$K = \sum_{s=1}^{L} \mathrm{pdf}(s)s^2, \qquad (16)$$

as the averaged squared cluster size. Here pdf($s$) is the probability density function to observe an event with the elongation $s$ in the sequence of avalanches.

Figure 7 presents the dependence of the susceptibility $K$ on the field parameter $t = 1/\alpha$ on log-log axes for different model sizes. The susceptibility $K$ increases monotonically with the decrease of $t$. Initially this increase is influenced by non-linear effects because the system is far from the fixed point of the renormalization group. When the field parameter $t$ reaches the vicinity of the fixed point, the linearization of the renormalization group becomes possible. Starting from this value of $t$ the dependence of the susceptibility on the field parameter becomes a power-law $1/t^\gamma$ with the exponent $\gamma = 2.94 \pm 0.03$. This value was obtained by the maximum likelihood fit of the power-law parts of the curves for the SBMs with 500 and 1000 blocks. We use for the fit only these model sizes because they provide the dependence which is the cleanest from the non-linear and crossover effects. For the infinite system we would expect the power-law divergence of the susceptibility at the critical point $t = 0$ with the same value of the exponent $\gamma$.

However, our SBMs are finite. Therefore, when the system approaches the critical point and the correlation length reaches the size of the system, the susceptibility stops to increase as a power-law and stays constant. Starting from this value of the field parameter, the finite-size effect, as a crossover effect, influences the dependence of $K$ on $t$. In other words, the mean cluster size reaches the limit of the system size.

From the theory of scaling functions [9, 11, 17] we expect that the functional dependence of the susceptibility $K$ on the field parameter $t$ should have the form

$$K \propto \frac{1}{t^\gamma} \Xi\left(Lt^\nu\right), \text{ where } \Xi(x) = \begin{cases} const, \, x \gg 1 \\ x^{\gamma/\nu}, \, x \ll 1 \end{cases} \qquad (17)$$

is some scaling function. In the limit $Lt^\nu \gg 1$, when $\xi \ll L$, this function has a constant limit, which does not influence the power-law dependence $1/t^\gamma$. In the limit $Lt^\nu \ll 1$, when the correlation length of the infinite system is $\xi_\infty \gg L$, this function generates a power-law dependence $x^{\gamma/\nu} \propto t^\gamma$, which cancels the power-law dependence $1/t^\gamma$ in front of the function $\Xi(x)$ and provides the finite limit for the susceptibility. To find the scaling function $\Xi(x)$ we multiply the dependence $K(t)$ by $t^\gamma$ and then plot the resulting dependence $Kt^\gamma$ as a function of the parameter $x = Lt^\nu$. The obtained scaling function $\Xi(x)$ is presented in figure 8. Also we plot the dependence $x^{\gamma/\nu}$ to compare it with the scaling function for low values of $x$. In figure 8 we see that all curves perfectly collapse on the scaling dependence $\Xi(x)$ except only for high values of $t$ when the system is far from the critical point and the renormalization group cannot be linearized far from its fixed point. At these high values of *t* the power-law divergence $1/t^{\gamma}$ of the susceptibility has non-linear deviations, and the scaling is not valid.

## 7. Frequency-size distribution

In section 3 we discussed the frequency-size behavior of the SBM. In this section we return to the frequency-size distribution to investigate its scaling.

From the theory of scaling functions [9, 11, 17] we expect that the functional dependence of the frequency-size distribution on the field parameter *t* and on the size *s* of an event should have the form

$$FSD \propto \frac{1}{s^{\tau}} \Xi\left(st^{\nu}, Lt^{\nu}\right) \qquad (17)$$

where $\tau$ is the scaling exponent, discussed in Section 3. Further on we will use $\tau = 2$. To find the scaling function $\Xi(x,y)$ we should multiply the *FSD* dependence by $s^{\tau}$ and then plot the resulting dependence $FSDs^{\tau}$ as a function of the parameters $x = st^{\nu}$ and $y = Lt^{\nu}$. However, in contrast to other scaling dependences discussed above, we encounter here a difficulty. If we were looking at a percolating system, the frequency-size distribution would be normalized by the size of the lattice. In other words, the number of possibilities to count a particular cluster configuration is limited by the lattice size, which gives a natural normalization for the distribution. In the case of the SBM we count clusters as they occur in time during the model evolution. The time of possible observations is not limited, and in our model we lost a natural normalization of the frequency-size distribution. Therefore, observing the scaling function $\Xi(x,y)$, we can determine it only with the accuracy of a constant multiplier. In figure 9(a,b) we plot the obtained scaling dependences on the log-log-log axes for all five model sizes $L = 25$, 50, 100, 500, and 1000, each above other. All obtained scaling functions $\Xi(x,y)$ have similar shapes and similar tendencies to become straight horizontal lines when the stiffness of the model increases.

## 8. Correlation function

Following the theory of percolation [13], we define the correlation function $G(\vec{R})$ as a probability that, if a given site is occupied, the site at distance $\vec{R}$ is also occupied and belongs to the same cluster

$$G(\vec{R}) = \frac{\sum_{k} \sum_{<i,j=i+\vec{R}> \in \text{ cluster } k} 1}{\sum_{k} \sum_{i \in \text{ cluster } k} 1}, \tag{18}$$

where the sum $\sum_{k}$ goes over all clusters enumerated by the index $k$, the sum $\sum_{i \in \text{ cluster } k}$ goes over all sites of cluster $k$, and the sum $\sum_{<i,j=i+\vec{R}> \in \text{ cluster } k}$ goes over all pairs of sites $<i,j>$ which belong to cluster $k$ and which are separated by the distance $\vec{R}$. Arranging clusters by their size we obtain

$$G(\vec{R}) = \frac{\sum_s \sum_{k_s=1}^{N_s} \sum_{<i,j=i+\vec{R}> \in \text{ cluster } k_s} 1}{\sum_s \sum_{k_s=1}^{N_s} s}, \tag{19}$$

where index $k_s$ enumerates $N_s$ clusters of size $s$. For the one-dimensional SBM, when clusters are linear chains, we can significantly simplify equation (19) as

$$G(R) = \frac{\sum_s pdf(s) \left\{ \begin{array}{l} s, R = 0 \\ 0, R > 0, s < R+1 \\ s-R, R > 0, s \geq R+1 \end{array} \right\}}{\sum_s pdf(s)s}, \tag{20}$$

The behavior of the correlation function $G(R)$ as a function of distance $R$ for different stiffnesses of the model of size $L = 1000$ is presented in Fig. (10a) on log-log axes. The observation we can make is that for the power-law part of the dependence, which on the log-log axes is supposed to be a straight line, we observe zero exponent, in other words, a horizontal line. Therefore, for the dependence $G(R) \propto \Xi(-R/\xi)/R^\eta$, where $\Xi(x)$ is some near-exponential function, we expect the exponent $\eta$ to be zero. This is confirmed by Fig. (10b), where we plot the correlation function on the semi-log axes. The main attenuating dependence is near exponential without a power-law addition. However, it is not pure exponential and for high $R$ has attenuation faster than exponential.

From the theory of scaling functions [9, 11, 17] we expect that the functional dependence of the correlation function on the field parameter $t$ and on the distance $R$ should have the form

$$G \propto \frac{1}{R^{\eta}} \Xi\left(Rt^{\nu}, Lt^{\nu}\right) \propto \Xi\left(Rt^{\nu}, Lt^{\nu}\right) \text{ when } \eta = 0. \qquad (21)$$

In Fig. (10c) we plot the correlation function as a function of the parameters $x = Rt^{\nu}$ and $y = Lt^{\nu}$ on the log-log-log axes for all five model sizes $L = 25$, 50, 100, 500, and 1000. All obtained scaling functions $\Xi(x,y)$ collapse on a single surface with minor non-linear deviations far from the critical point where the renormalization group cannot be linearized.

## 9. Conclusions

For different sizes of the slider-block model we obtain the dependence of the correlation length on the stiffness of the system as a field parameter. The obtained scaling suggests that the slider-block model has a critical point when its stiffness is infinite. For the exponents of the correlation length and susceptibility we obtain values 1.85 and 2.94 respectively. Also we investigate the finite-size scaling functions of the model and find that the dependence for different model sizes collapses onto a single curve. For the exponents of the frequency-size distribution and correlation function we find $\tau = 2$ and $\eta = 0$ respectively.

**References**

[1]     Burridge R and Knopoff L, *Model and theoretical seismicity*, 1967 *Bull. Seism. Soc. Am.* **57** 341

[2]     Carlson J M and Langer J S, *Mechanical model of an earthquake fault*, 1989 *Phys. Rev. A* **40** 6470

[3]     Abaimov S G, Turcotte D L, Shcherbakov R and Rundle J B, *Recurrence and interoccurrence behavior of self-organized complex phenomena*, 2007 *Nonlinear Proc. Geophys.* **14** 455

[4]     Abaimov S G, Turcotte D L, Shcherbakov R, Rundle J B, Yakovlev G, Goltz C and Newman W I, *Earthquakes: Recurrence and interoccurrence times*, 2008 *Pure Appl. Geophys.* **165** 777

[5]     Abaimov S G, Tiampo K F, Turcotte D L and Rundle J B, *Recurrent frequency-size distribution of characteristic events*, 2009 *Nonlinear Proc. Geophys.* **16** 333

[6]     Drossel B and Schwabl F, *Self-organized critical forest-fire model*, 1992 *Phys. Rev. Lett.* **69** 1629

[7]     Grassberger P, *Critical behaviour of the Drossel-Schwabl forest fire model*, 2002 *New J. Phys.* **4** 17

[8]     Bak P, Tang C and Wiesenfeld K, *Self-organized criticality*, 1988 *Phys. Rev. A* **38** 364

[9]  Goldenfeld N, 1992 *Lectures on Phase Transitions and the Renormalization Group* (Reading, MA: Addison Wesley)

[10]  Cardy J, 1996 *Scaling and Renormalization in Statistical Physics* (Cambridge: Cambridge University Press)

[11]  Pathria R K, 1996 *Statistical Mechanics* (Oxford: Butterworth-Heinemann)

[12]  Ma S K, 1976 *Modern Theory of Critical Phenomena* (Reading, MA: Benjamin)

[13]  Stauffer D and Aharony A, 1992 *Introduction To Percolation Theory*: Taylor&Francis)

[14]  Abaimov S G, *Applicability and non-applicability of equilibrium statistical mechanics to non-thermal damage phenomena*, 2008 *J. Stat. Mech.* P09005

[15]  Abaimov S G, *Applicability and non-applicability of equilibrium statistical mechanics to non-thermal damage phenomena: II. Spinodal behavior*, 2008 *J. Stat. Mech.* P03039

[16]  Thijssen J M, 1999 *Computational Physics* (Cambridge: Cambridge University Press)

[17]   Brankov J G, 1996 *Introduction to Finite-Size Scaling*, Leuven Notes in Mathematical and Theoretical Physics. Series A: Mathematical Physics vol 8 (Leuven: Leuven University Press)

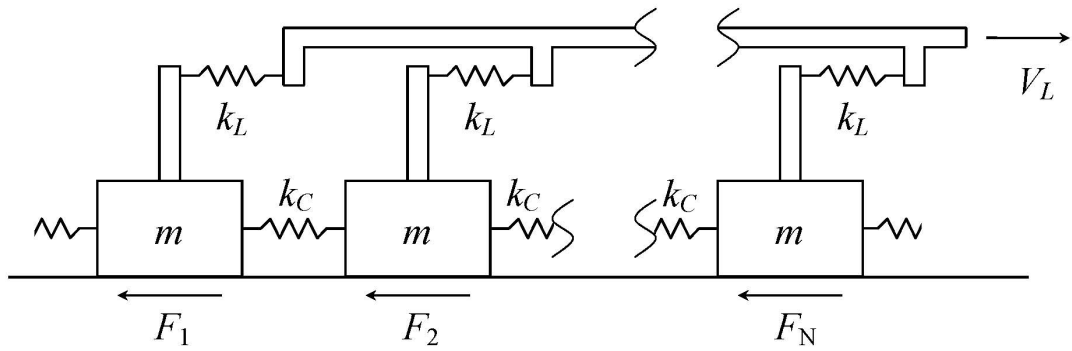

Figure 1. A slider-block model.

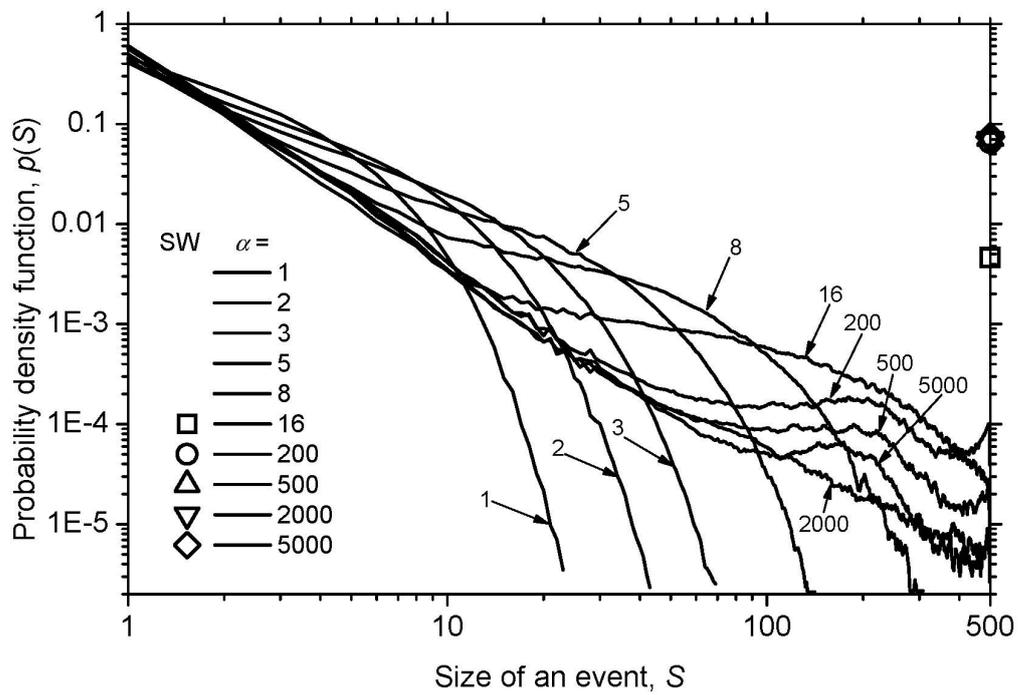

(a)

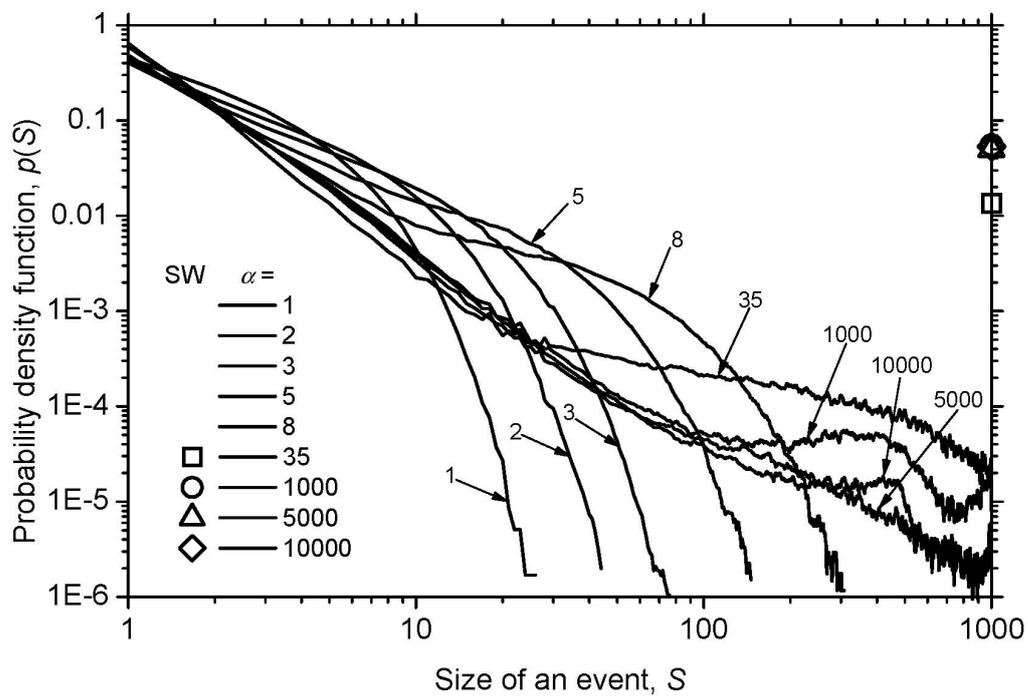

(b)

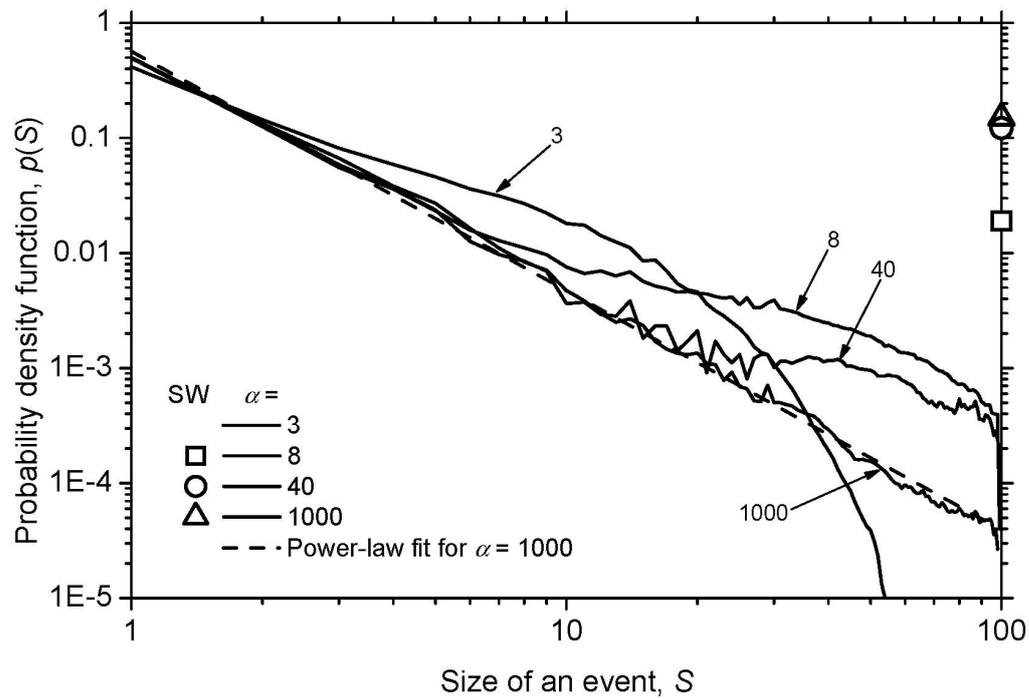

(c)

Figure 2. Frequency-size distribution of the model with (a) $L = 500$ blocks, (b) $L = 1000$ blocks, and (c) $L = 100$ blocks for different values of the model stiffness $\alpha$. The values of $\alpha$ are shown in the legends and in the labels for individual curves. Starting from (a) $\alpha = 16$, (b) $\alpha = 35$, and (c) $\alpha = 8$, system-wide (SW) events are shown as markers on the right sides of the plots.

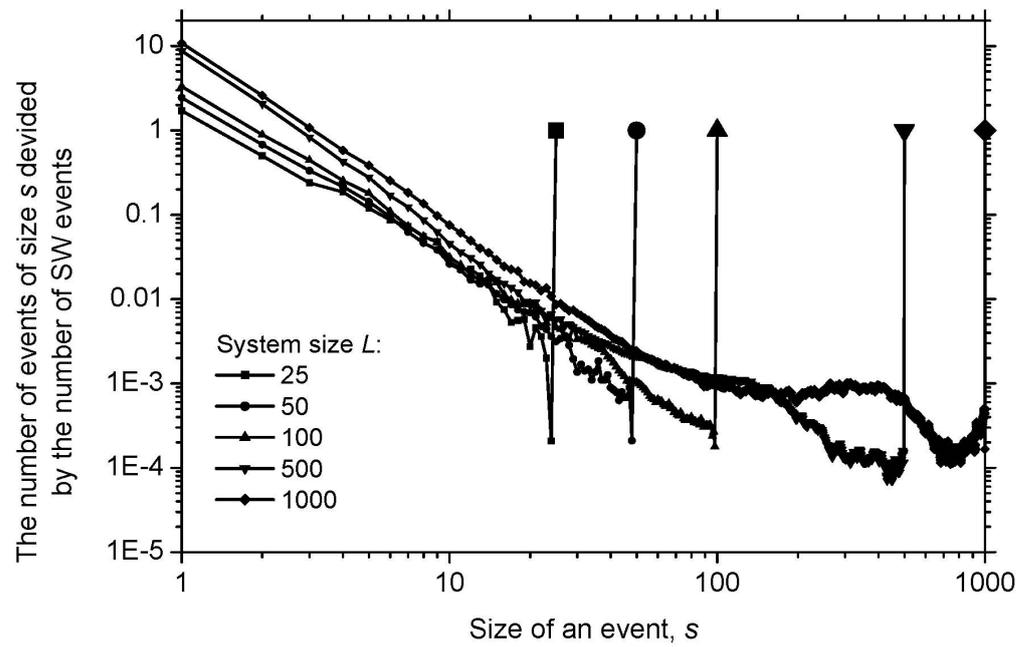

Figure 3. The frequency-size distribution normalized by the number of SW events. For all model sizes $L$ the value of $\alpha$ is 1000.

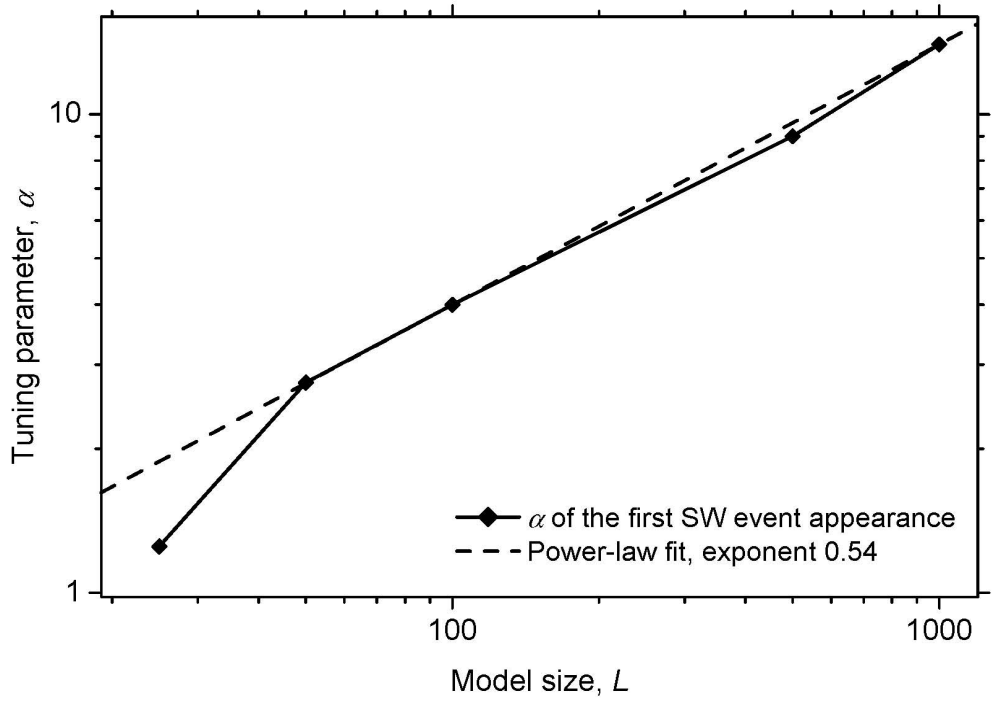

Figure 4. The values of $\alpha$, at which the first SW events appear. The fit shows that the dependence is close to the square root of the size of the model $L$.

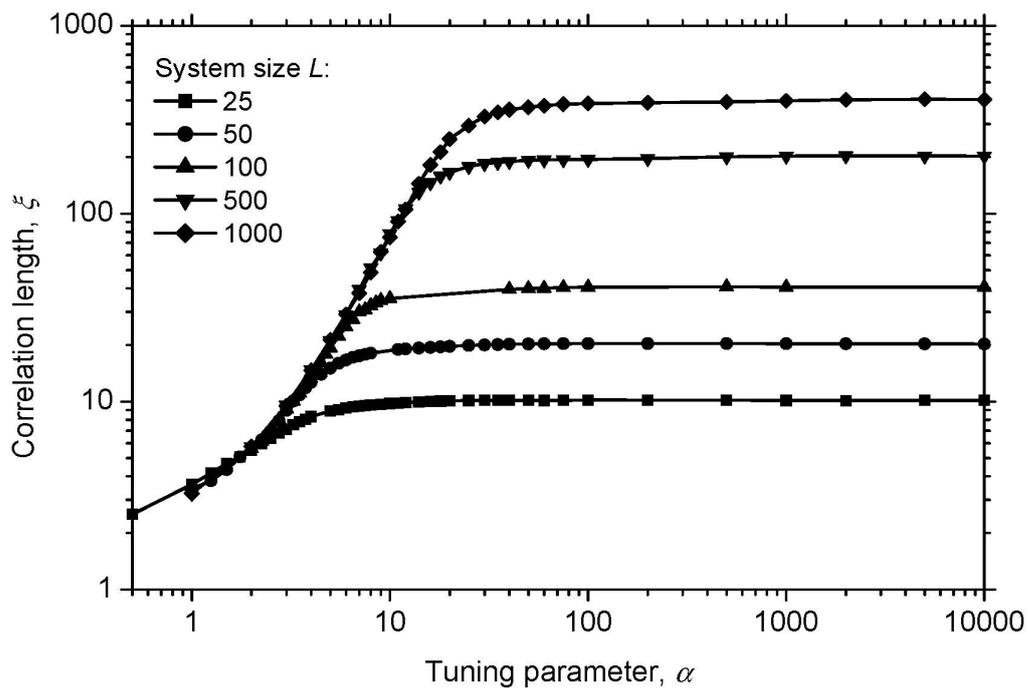

(a)

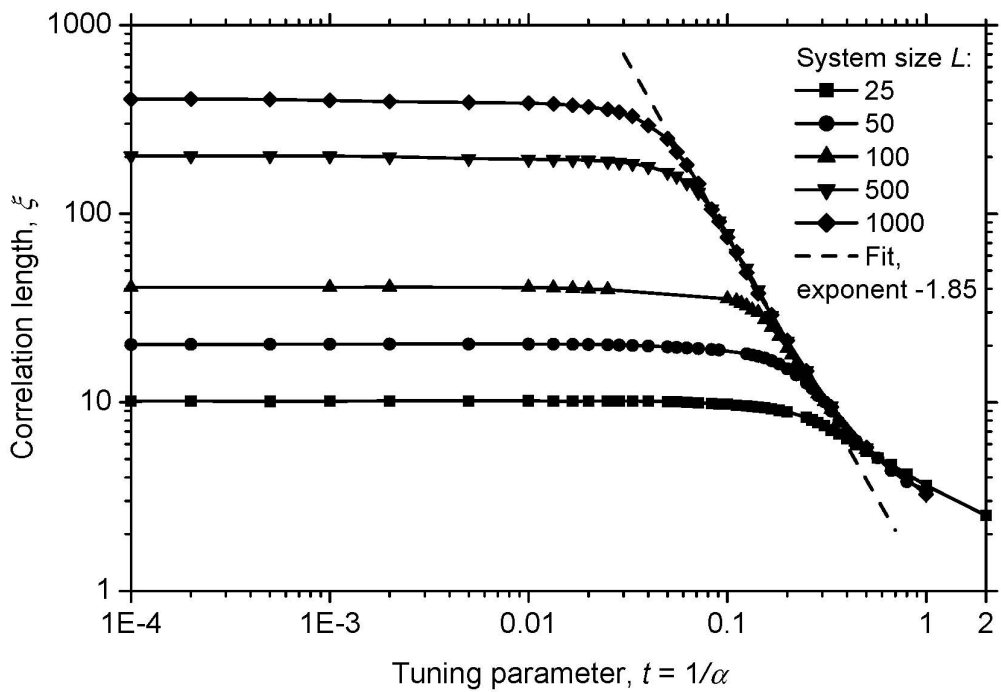

(b)

Figure 5. Correlation length $\xi$ as a function of the (a) field parameter $\alpha$ and (b) field parameter $t = 1 / \alpha$. Each marker represents a separate sequence of avalanches obtained in numerical simulations. The dashed line is the maximum likelihood fit for the power-law parts of the curves for the SBMs with 500 and 1000 blocks.

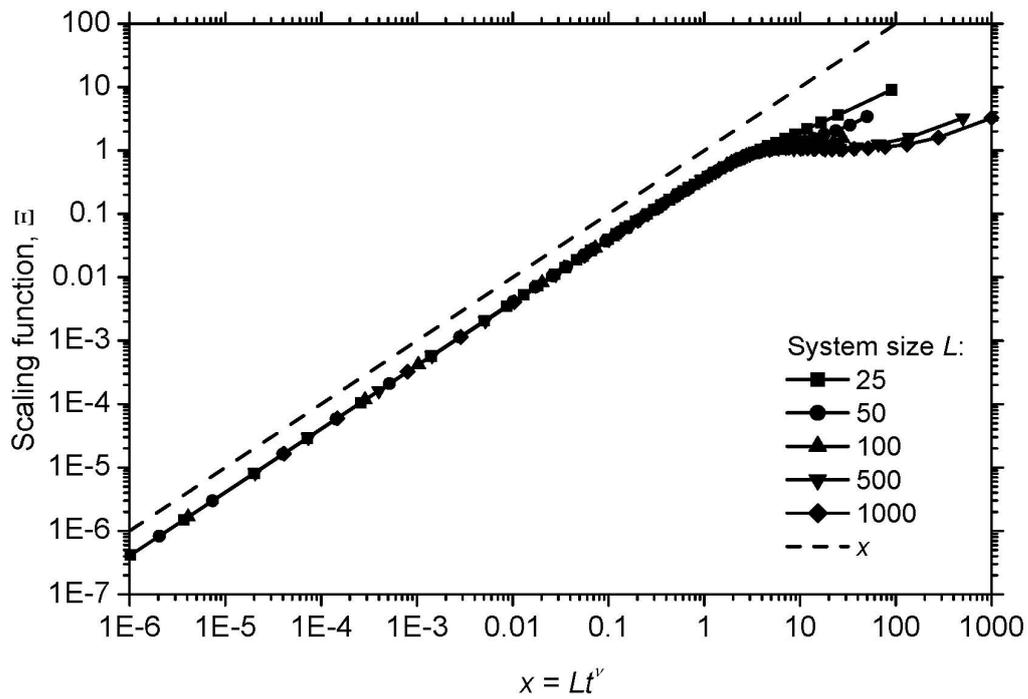

Figure 6. Scaling function $\Xi(x)$ of the correlation length $\xi$. For comparison, the dependence $x$ is given as a dashed line.

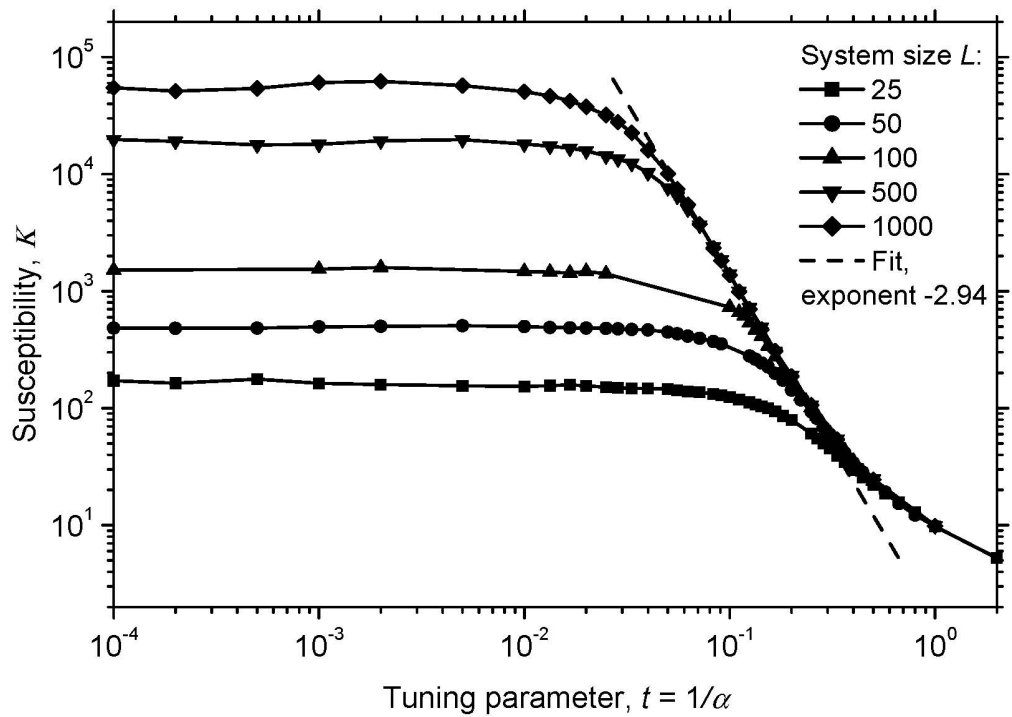

Figure 7. Susceptibility $K$ as a function of the field parameter $t = 1 / \alpha$. Each marker represents a separate sequence of avalanches obtained in numerical simulations. The dashed line is the maximum likelihood fit for the power-law parts of the curves for the SBMs with 500 and 1000 blocks.

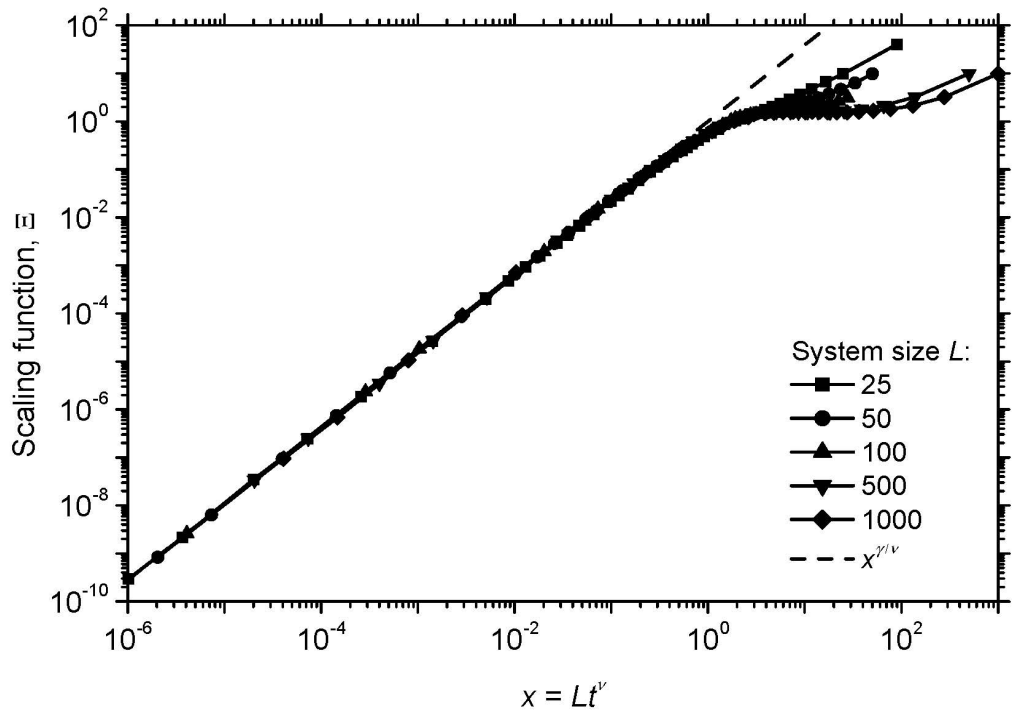

Figure 8. Scaling function $\Xi(x)$ of the susceptibility $K$. For comparison, the dependence $x^{\gamma/\nu}$ is given as a dashed line.

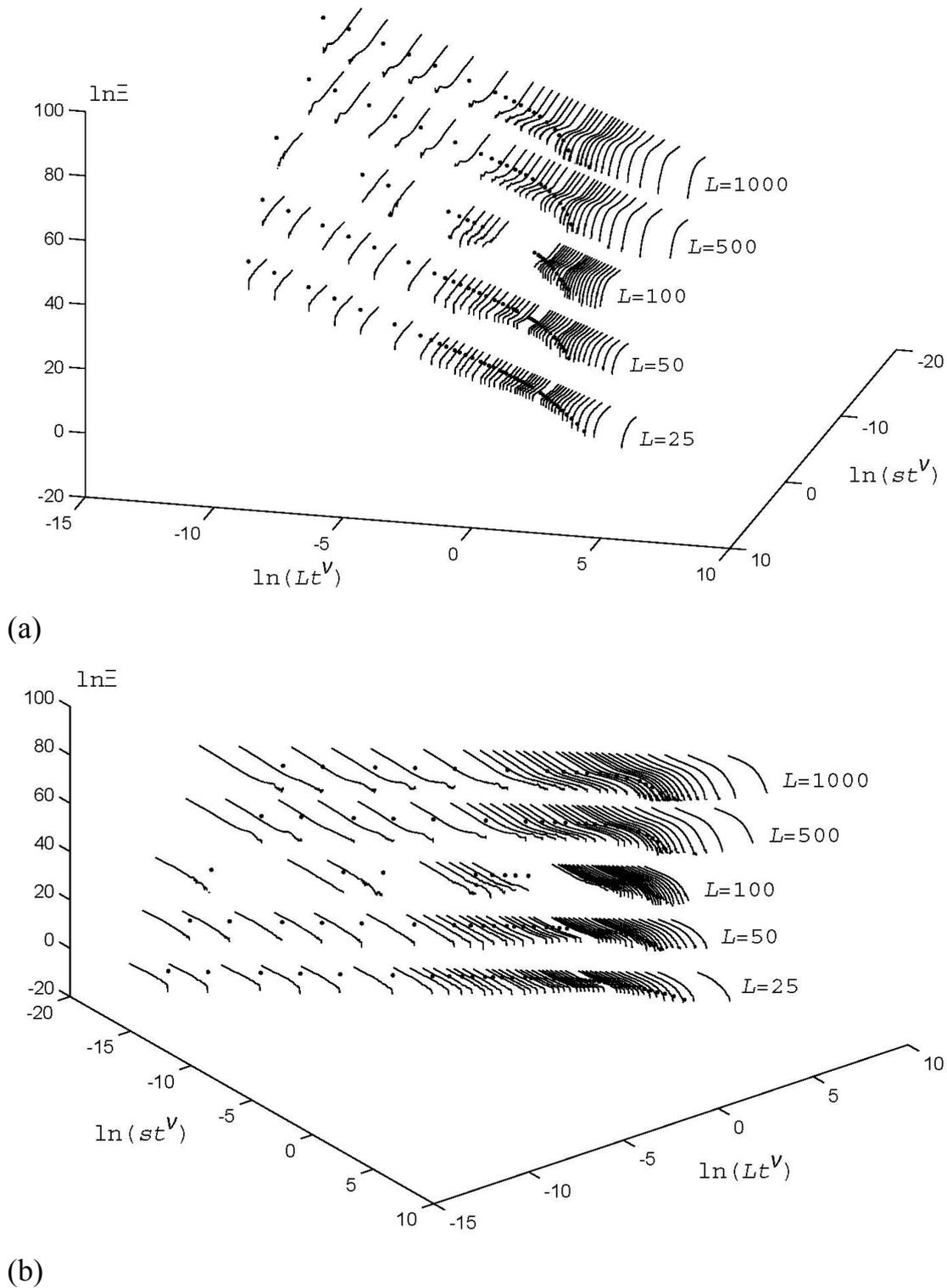

(a)

(b)

Figure 9. Scaling functions $\Xi(x)$ of the frequency-size distributions. Each distribution is labeled by its model size $L$. Each solid line represents a particular distribution over event sizes $s$ for given values of the field parameter $t$ and model size $L$. Horizontal shift among solid lines corresponds to the change of the field parameter $t$; vertical shift corresponds to the change of the model size $L$. Dot-markers represent SW events.

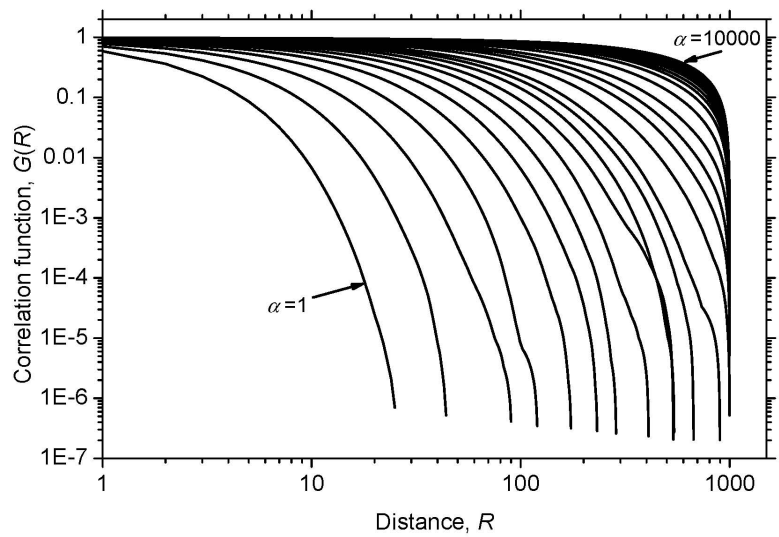

(a)

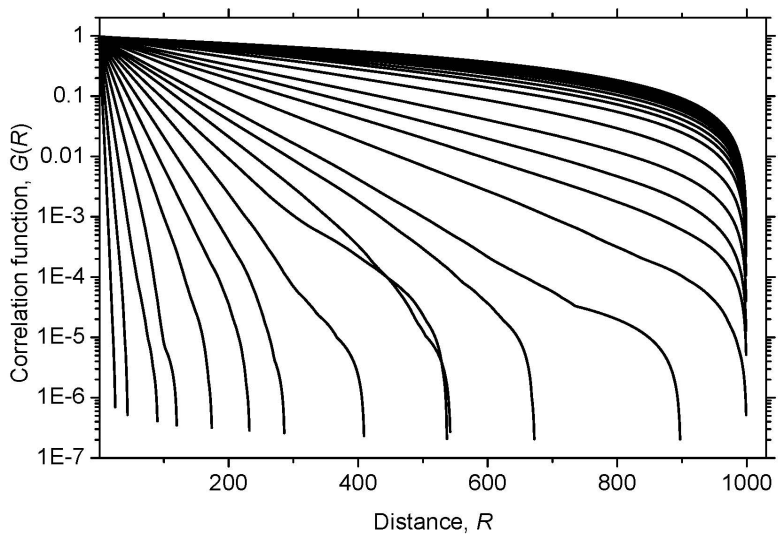

(b)

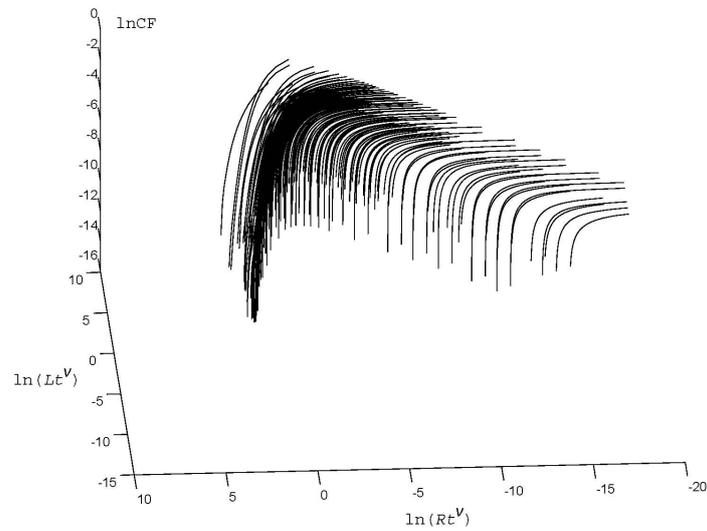

(c)

Figure 10. Correlation function. (a-b) For model size $L = 1000$ the dependence of the correlation function on distance $R$ is presented for different model stiffnesses on (a) log-log and (b) semi-log axes. (c) The scaling function of the correlation function. Each solid line represents a particular correlation function over distances $R$ for given values of the field parameter $t$ and model size $L$. Horizontal shift among solid lines corresponds to the change of the field parameter $t$. All model sizes are collapsed on a single surface.